\documentclass[10pt,conference]{IEEEtran}
\usepackage{amsmath,amssymb}
\usepackage{stmaryrd}
\usepackage[latin1]{inputenc}
\usepackage[numbers, square, comma, compress]{natbib}
\usepackage{dsfont}
\usepackage{graphicx}
\usepackage{epstopdf}
\usepackage[english]{babel}

\newcommand{\Q}{\mathbb{Q}}
\newcommand{\C}{\mathbb{C}}
\newcommand{\Y}{\mathsf{Y}}
\newcommand{\Z}{\mathsf{Z}}

\newcommand{\Hh}{\mathsf{H}}
\newcommand{\X}{\mathsf{X}}
\newcommand{\M}{\mathsf{M}}
\newcommand{\R}{\mathbb{R}}

\newcommand{\I}{\mathbb{I}}

\providecommand{\mb}[1]{\ensuremath{\mathbf #1}}

\providecommand{\abs}[1]{\ensuremath{\left\lvert #1 \right\rvert}}

\providecommand{\norm}[1]{\ensuremath{\left\Vert #1 \right\Vert}}

\providecommand{\vv}[1]{\textquotedblleft #1\textquotedblright}

\DeclareMathOperator*{\Vol}{Vol}
\DeclareMathOperator*{\diag}{diag}
\DeclareMathOperator*{\Tr}{Tr}
\DeclareMathOperator*{\argmax}{argmax}
\DeclareMathOperator*{\argmin}{argmin}

\newtheorem{thm}{Theorem}[section]
\newtheorem{theorem}[thm]{Theorem}

\newtheorem{lem}{Lemma}
\newtheorem{prop}{Proposition}

\newtheorem{rem}{Remark}
\newtheorem{defn}{Definition}

\date{}
\author{}

\begin{document}
\title{Almost universal codes for fading wiretap channels}

\author{
\IEEEauthorblockN{Laura Luzzi}
\IEEEauthorblockA{Laboratoire ETIS \\
 (ENSEA - UCP - CNRS) \\
Cergy-Pontoise, France \\
laura.luzzi@ensea.fr}
\and
\IEEEauthorblockN{Cong Ling}
\IEEEauthorblockA{Department of Electrical \\ and Electronic Engineering \\
Imperial College London, U.K.\\
c.ling@imperial.ac.uk}
\and
\IEEEauthorblockN{Roope Vehkalahti}
\IEEEauthorblockA{Department of Mathematics \\ and Statistics,\\
University of Turku, Finland\\
roiive@utu.fi}
}
\maketitle

\begin{abstract}
We consider a fading wiretap channel model where the transmitter has only statistical channel state information, and the legitimate receiver and eavesdropper have perfect channel state information. We propose a sequence of non-random lattice codes which achieve strong secrecy and semantic security over
ergodic fading channels. The construction is almost universal in the sense that it achieves the same constant gap to secrecy capacity over Gaussian and ergodic fading models. 
\end{abstract}

\section{Introduction}
The wiretap channel model was introduced by Wyner \cite{Wyner}, who showed that secure and reliable communication can be achieved simultaneously over noisy channels even without the use of secret keys. In the information theory community, the most widely accepted secrecy metric is Csiszár's \emph{strong secrecy}: the mutual information $\I(\M;\Z^n)$ between the confidential message $\M$ and the channel output $\Z^n$ should vanish when the code length $n$ tends to infinity. 

While in the information theory community confidential messages are often assumed to be uniformly distributed, this assumption is not accepted in cryptography. A cryptographic treatment of the wiretap channel was proposed in \cite{Bellare_Tessaro_Vardy} to combine the requirements of the two communities, establishing that achieving \emph{semantic security} in the cryptographic sense is equivalent to achieving strong secrecy for all distributions of the message. This equivalence holds also for continuous channels \cite{LLBS}.

In the case of Gaussian wiretap channels, \cite{LLBS} considered the problem of designing lattice codes which achieve strong secrecy and semantic security. Following an approach by Csiszár \cite{Csiszar, Bloch_Laneman}, 
strong secrecy is guaranteed if the output distributions of the eavesdropper's channel corresponding to two different messages are indistinguishable in the sense of variational distance. Moreover, the \emph{flatness factor} of a lattice was proposed in \cite{LLBS} as a fundamental criterion which implies that conditional outputs are indistinguishable. Using random lattice coding arguments, it was shown that there exist families of lattice codes which are \vv{good for secrecy}, meaning that their flatness factor is vanishing, and achieve semantic security for rates up to $1/2$ nat from the secrecy capacity.

In this paper, we consider a fading wiretap channel model where the transmitter has only access to statistical channel state information (CSI), while the legitimate receiver and the eavesdropper both have perfect knowledge of their own channels. We extend the criterion based on the flatness factor to the case of fading channels
and propose a family of non-random lattice codes from algebraic number fields satisfying this criterion. 
We note that ideal lattices from number fields were already considered for secrecy under an error probability criterion for Gaussian and fading channels in \cite{BO_ICC,B0_TComm,KHHV,Ong_Oggier} and related works. 

In this paper, we consider a particular sequence of algebraic number fields with constant root discriminant. In \cite{ISIT2015_SISO, LV}, it was shown that these lattice codes are \vv{almost universal} in the sense that they achieve a constant gap to channel capacity over \emph{any} ergodic stationary fading channel. The underlying multiplicative structure and constant root discriminant property guarantee that the received lattice after fading has a good minimum distance when the channel is not in outage.

The sequences of number fields that we consider are also used in the crypto literature  
for worst-case to average-case reductions of hard lattice problems  \cite{Peikert_Rosen}.
 
In this work, we show that these lattices also achieve 
strong secrecy and semantic security. 
The key feature is that the \emph{dual} of the faded lattice has good minimum distance, so that the average flatness factor of the faded lattice vanishes. 

In particular, for the Gaussian case this suggests a simple design criterion where the packing density of the lattice and its dual should be maximized simultaneously. 
We note that the dual code also plays a role in the design of 
LDPC codes for  binary erasure wiretap channels \cite{Subramanian}.

We also improve the rate of almost universal codes by replacing spherical shaping with a discrete Gaussian distribution over the infinite lattice as in \cite{LLBS}. As a consequence, our nested lattice schemes achieve the same constant gap to secrecy capacity over all static and ergodic fading models.

The proposed lattice codes can be generalized in a straightforward manner to the multi-antenna case using the multiblock matrix lattices from division algebras in \cite{LV}. This generalization 
will be presented in an upcoming journal version.

\section{Preliminaries}

\subsection{Flatness factor and discrete Gaussian distribution}
In this section, we define some fundamental lattice parameters that will be used in the rest of the paper. For more background about the smoothing parameter and the flatness factor in information theory and cryptography, we refer the reader to \cite{Micciancio_Regev,LLBS, Peikert}. \\
Consider $\C^{k}$ as a $2k$-dimensional real vector space with a real inner product $\langle \mb{x},\mb{y}\rangle=\Re(\mb{x}^{\dagger}\mb{y})$. This inner product naturally defines a metric on 
 $\C^k$ by setting $\norm{\mb{x}}= \sqrt{\langle \mb{x},\mb{x}\rangle}$. \footnote{This inner product corresponds to identifying $\C^k$ with $\R^{2k}$ with the canonical real inner product, through the isometry $\phi(z_1,\ldots,z_k)=(\Re(z_1),\ldots,\Re(z_k),\Im(z_k),\ldots,\Im(z_k))$. Note also that if $\Sigma=\Sigma^{\dagger}$, then $\langle \mb{z}, \Sigma \mb{z} \rangle=\Re(\mb{z}^{\dagger} \Sigma \mb{z})=\mb{z}^{\dagger} \Sigma \mb{z}=\phi(\mb{z})^T \Sigma_{\R} \phi(\mb{z})$, where
$\Sigma_{\R}=\begin{pmatrix} \Re(\Sigma) & -\Im(\Sigma) \\ \Im(\Sigma) & \Re(\Sigma)\end{pmatrix}$. In particular, the properties of real Gaussian distributions carry over to circularly symmetric complex Gaussian distributions.} \\
Given a complex lattice $\Lambda \subset \C^k$, we define the dual lattice as 
$$
\Lambda^*=\{\mb{x} \in \C^k \;|\; \forall \mb{y} \in \Lambda , \;\; \langle \mb{x},\mb{y} \rangle \in \mathbb{Z}\}.
$$
Let $f_{\sqrt{\Sigma},\mathbf{c}}(\mathbf{z})$ denote the $k$-dimensional complex normal distribution with mean $\mathbf{c}$ and covariance matrix $\Sigma$:
$$f_{\sqrt{\Sigma},\mathbf{c}}(\mathbf{z})=\frac{1}{\pi^k \det(\Sigma)} e^{-(\mb{z}-\mb{c})^{\dagger}\Sigma^{-1} (\mb{z}-\mb{c})} \quad \forall \mb{z} \in \C^k.$$
We will use the notation $f_{\sigma,\mb{c}}(\mb{z})$ for $f_{\sigma I,\mb{c}}(\mb{z})$. 
\begin{defn}
Given a complex lattice $\Lambda \subset \C^k$, the \emph{flatness factor} $\epsilon_{\Lambda}(\sqrt{\Sigma})$ is defined as the maximum deviation of the Gaussian distribution over $\Lambda$ from the uniform distribution over a fundamental region $\mathcal{R}(\Lambda)$ of $\Lambda$, with volume $V(\Lambda)$:
$$\epsilon_{\Lambda}(\sqrt{\Sigma})=\max_{\mathbf{z} \in \mathcal{R}(\Lambda)} \abs{V(\Lambda)\sum_{\boldsymbol\lambda \in \Lambda}f_{\sqrt{\Sigma},\lambda}(\mathbf{z})-1}.$$
\end{defn}
Compared to \cite{LLBS}, in this paper we use an extended version of the flatness factor for correlated Gaussians, related to the extended notion of the smoothing parameter in \cite{Peikert}. We also extend the definition to the case of complex lattices.\\
Note that correlations can be absorbed by the lattice in the sense that $\epsilon_{\Lambda}(\sqrt{\Sigma})=\epsilon_{\sqrt{\Sigma}^{-1}\Lambda}(I)$, and that $\epsilon_{\Lambda}(\sqrt{\Sigma_1}) \leq \epsilon_{\Lambda}(\sqrt{\Sigma_2})$ if $\Sigma_1$ and $\Sigma_2$ are two positive definite matrices with $\Sigma_1 \succeq \Sigma_2$. 
\begin{defn}
Given a lattice $\Lambda$ and $\epsilon>0$, the \emph{smoothing parameter}\footnote{Note that we define the smoothing parameter per complex dimension, which differs by a factor $\sqrt{2}$ from the definition in \cite{Micciancio_Regev}. We have adjusted the bounds on $\eta_{\epsilon}(\Lambda)$ accordingly.} $\eta_{\epsilon}(\Lambda)$ is the smallest $s=\sqrt{2\pi}\sigma>0$ such that $\sum_{\boldsymbol\lambda^* \in \Lambda^*\setminus \{\mb{0}\}} 
e^{-\pi^2 \sigma^2 \norm{\boldsymbol\lambda^*}^2} \leq \epsilon$, where $\Lambda^*$ is the dual lattice.
\end{defn}
To extend the definition to matrices we can say that 
\begin{equation} \label{smoothing_parameter_relation}
 \sqrt{2\pi\Sigma} \succeq \eta_{\epsilon}(\Lambda) \quad \text{if} \quad \epsilon_{\Lambda}(\Sigma) \leq \epsilon.
\end{equation}
The smoothing parameter is upper bounded by the minimum distance of the dual lattice \cite{Micciancio_Regev}:
\begin{equation} \label{Micciancio_Regev_dual}
 \eta_{\epsilon}(\Lambda) \leq \frac{2\sqrt{k}}{\lambda_1(\Lambda^*)}.
 \end{equation}
Finally, given $\mb{c} \in \C^k$ and $\sigma>0$, we define the \emph{discrete Gaussian distribution} over the (shifted) lattice $\Lambda - \mb{c} \subset \C^{k}$ as the following 
discrete distribution taking values in $\Lambda - \mb{c}$: 
$$D_{\Lambda-\mb{c},\sigma}(\boldsymbol\lambda-\mb{c})=\frac{f_{\sigma}(\boldsymbol{\lambda}-\mb{c})}{f_{{\sigma},\mb{c}}(\Lambda)}.$$

The following result is a consequence of \cite[Theorem 3.1]{Peikert} and extends Lemma 8 in \cite{LLBS}:
\begin{lem} \label{extended_Regev_Lemma}
Let $\X_1$ be sampled according to the discrete Gaussian distribution $D_{\Lambda + \mb{c},\sqrt{\Sigma_1}}$ and $\X_2$ be sampled according to the continuous Gaussian $f_{\sqrt{\Sigma_2}}$. Let $\Sigma_0=\Sigma_1+\Sigma_2$ and $\Sigma^{-1}=\Sigma_1^{-1}+\Sigma_2^{-1}$. If 
\begin{equation} \label{epsilon_condition}
\epsilon_{\Lambda}(\sqrt{\Sigma}) \leq \epsilon \leq \frac{1}{2},
\end{equation}
then the distribution $g$ of $\X=\X_1+\X_2$ is close to $f_{\sqrt{\Sigma_0}}$: 
$$\mathbb{V}(g,f_{\sqrt{\Sigma_0}})\leq 4 \epsilon,$$
where $\mathbb{V}(\,,\,)$ is the $L^1$ distance. 
\end{lem}

\subsection{Ideal lattices from number fields with constant root discriminant}

Let $F$ be a number field of degree $[F:\Q]=n$, with ring of integers $\mathcal{O}_F$. We denote by $d_F$ the discriminant of the number field.
We define the \emph{codifferent} of $F$ as
$$\mathcal{O}_F^{\vee}=\{ x \in K: {\Tr}_{F/\Q}(x\mathcal{O}_F) \subseteq \mathbb{Z}\}.$$   
The codifferent is a fractional ideal, that is, there exists some integer $a$ such that $a\mathcal{O}_F^{\vee}$ is a proper ideal of $\mathcal{O}_F$, and its algebraic norm is the inverse of the discriminant:
\begin{equation} \label{discriminant}
N(\mathcal{O}_F^{\vee})=1/d_F.
\end{equation}
We will focus on the case of totally complex extensions $F/\Q$ of degree $n=2k$. The \emph{relative canonical embedding} of $F$ into $\C^k$ is given by
$$\psi(x)=(\sigma_1(x),\ldots,\sigma_k(x)),$$
where $\{\sigma_1,\ldots,\sigma_k\}$ is a set of $\Q$-embeddings $F \to \C$ such that we have chosen one from each complex conjugate pair. \\
Then $\Lambda=\psi(\mathcal{O}_F)$ is a lattice in $\C^k$. The codifferent embeds as the complex conjugate of the dual lattice: 
\begin{equation} \label{dual_lattice}
\Lambda^*=2\overline{\psi(\mathcal{O}_F^{\vee})}.
\end{equation}
Using (\ref{Micciancio_Regev_dual}), we obtain 
\begin{equation} \label{dual}
\eta_{\epsilon}(\Lambda) \leq \frac{\sqrt{k}}{\lambda_1(\overline{\psi(\mathcal{O}_F^{\vee})})}. 
\end{equation}
From the AM-GM inequality we have that for any fractional ideal $\mathcal{I}$ of $\mathcal{O}_F$,
$$\lambda_1(\psi(\mathcal{I})) \geq \sqrt{k} (N(\mathcal{I}))^{\frac{1}{2k}}.$$ 
In particular, from (\ref{discriminant}) we get  
\begin{equation} \label{lambda1} 
\lambda_1(\overline{\psi(\mathcal{O}_F^{\vee})})=\lambda_1(\psi(\mathcal{O}_F^{\vee}))\geq \frac{\sqrt{k}}{\abs{d_F}^{\frac{1}{2k}}}.
\end{equation}
Combining equations (\ref{dual}) and (\ref{lambda1}), we find that the smoothing parameter of $\Lambda$ is upper bounded by the root discriminant \cite[Lemma 6.5]{Peikert_Rosen}: given $\epsilon=2^{-2k}$,
\begin{equation} \label{Peikert_Rosen_bound}
\eta_{\epsilon}(\Lambda) \leq \abs{d_F}^{\frac{1}{2k}}.
\end{equation}

The following theorem by Martinet \cite{Martinet} proves the existence of infinite towers of totally complex number fields with constant root discriminant.
\begin{theorem} \label{Martinet_theorem} 
There exists an infinite tower of totally complex number fields $\{F_k\}$ of degree $2k=5\cdot2^t$, such that
\begin{equation} \label{G}
 \abs{d_{F_k}}^{\frac{1}{2k}}=G,
\end{equation}
for $G \approx 92.368$.
\end{theorem}

We now focus on the corresponding lattice sequence $\Lambda^{(k)} \subset \C^{k}$. Their volume is a function of the discriminant:
\begin{equation} \label{volume_nf}
\Vol(\Lambda^{(k)})=2^{-k}\sqrt{\abs{d_F}}=2^{-k} G^k 
\end{equation}
Let $\epsilon=2^{-2k}$. From Theorem \ref{Martinet_theorem} and equation (\ref{Peikert_Rosen_bound}), 
$$\eta_{\epsilon}(\Lambda^{(k)})\leq \abs{d_F}^{\frac{1}{2k}}=G.$$ 
Since the flatness factor is a decreasing function of $\sigma$, 
\begin{equation} \label{flatness_factor_nf}
\forall \sigma> \frac{G}{\sqrt{2\pi}}, \quad \varepsilon_{\Lambda^{(k)}}(\sigma) \leq 2^{-2k}. 
\end{equation}

\section{Fading wiretap channel }
We consider an ergodic fading channel model where the outputs $\Y^k$ and $\Z^k$ at Bob and Eve's end are given by
\begin{equation} 
\begin{cases}
{\Y}_i={\Hh}_{b,i}\X_i + \mathsf{W}_{b,i},\\
{\Z}_i={\Hh}_{e,i}\X_i + \mathsf{W}_{e,i},
\end{cases} \quad i=1,\ldots,k
\end{equation}
where $\mathsf{W}_{b,i}$, $\mathsf{W}_{e,i}$ are i.i.d. complex Gaussian vectors with zero mean and variance 
$\sigma_b^2$, $\sigma_e^2$ per complex dimension. The input $\X^k$ satisfies the average power constraint 
\begin{equation} \label{power_constraint}
\frac{1}{k} \sum_{i=1}^{k} \abs{\X_i}^2 \leq P. 
\end{equation}
We suppose that 
$\Hh_{b,i}$, $\Hh_{e,i}$ are isotropically invariant channels such that the channel capacities $C_b$ and $C_e$ are well-defined and the weak law of large numbers holds: $\forall \delta>0$,
{\allowdisplaybreaks
\begin{align} 
&\lim_{k \to \infty} \mathbb{P}\left\{ \abs{\frac{1}{k} \sum_{i=1}^k \ln \left(1+\frac{P}{\sigma_b^2}\abs{h_{b,i}}^2\right) - C_b}>\delta\right\}=0, \label{LLN_Bob}\\
&\lim_{k \to \infty} \mathbb{P}\left\{ \abs{\frac{1}{k} \sum_{i=1}^k \ln \left(1+\frac{P}{\sigma_e^2}\abs{h_{e,i}}^2\right) - C_e}>\delta\right\}=0.\label{LLN_Eve}\end{align}}
All rates are expressed in nats per complex channel use. \\
We suppose that Alice has no instantaneous CSIT (apart from knowledge of channel statistics), and Bob and Eve have perfect CSI of their own channels.
A confidential message $\M$ and an auxiliary message $\M'$ with rate $R$ and $R'$ respectively are encoded into $\X^k$. We denote by $\hat{\M}$ the estimate of the confidential message at Bob's end. 

\begin{defn}
A coding scheme achieves \emph{strong secrecy} if
\begin{align*}
&\lim_{k \to \infty} \mathbb{P}\{\hat{\M} \neq \M\} = 0,  \quad &\text{(reliability condition)}\\
&\lim_{k \to \infty} \mathbb{I}(\M; \Z^k, \Hh_e^k)=0. \quad &\text{(secrecy condition)}
\end{align*} 
\end{defn}

The secrecy capacity for this wiretap model is given by \cite{Lin}
\begin{equation} \label{secrecy_capacity}
C_s=C_b-C_e.
\end{equation}

Let  $\Lambda^{(k)} \subset \C^{k}$ be the lattice sequence defined in the previous section. We consider scaled versions $\Lambda_b=\alpha_b \Lambda^{(k)}$, $\Lambda_e=\alpha_e \Lambda^{(k)}$ such that $\Lambda_e \subset \Lambda_b$ and $\abs{\Lambda_b/\Lambda_e}=e^{kR}$.

We consider the secrecy scheme in \cite{LLBS}, where each confidential message $m \in \mathcal{M}=\{1,\ldots,e^{kR}\}$ is associated to a 
coset leader $\boldsymbol{\lambda}_m \in \Lambda_b \cap \mathcal{R}(\Lambda_e)$ for a fundamental region $\mathcal{R}(\Lambda_e)$. To transmit the message $m$, Alice samples $\X^k$ from the discrete Gaussian $D_{\Lambda_e + \boldsymbol{\lambda}_m,\sigma_s}$ with $\sigma_s^2=P$. It follows from \cite[Lemma 6]{LLBS} that as $k \to \infty$, the variance per complex dimension of $\X^k$ tends to $P$ provided that 
\begin{equation} \label{Lemma6_condition}
\lim_{k \to \infty} \epsilon_{\Lambda_e}(\sqrt{P}) = 0.
\end{equation}
From \cite[Lemma 7]{LLBS}, the information rate $R'$ of the auxiliary message (corresponding to the choice of a point in $\Lambda_e$) is 
$$ R' \approx \ln(\pi eP)-\frac{1}{k} \ln V(\Lambda_e)=\ln(\pi e P) - \frac{1}{k} \ln(\alpha_e^{2k}2^{-k}G^{k}).$$  
Therefore, we have
\begin{equation} \label{alpha_e}
\alpha_e^2 \approx \frac{2\pi eP}{G e^{R'}}.  
\end{equation}
From (\ref{flatness_factor_nf}), $ \epsilon_{\Lambda_e}(\sqrt{P})=\epsilon_{\alpha_e\Lambda}(\sqrt{P}) =\epsilon_{\Lambda}\left(\sqrt{P}/\alpha_e\right) \to 0$
provided that $\frac{\sqrt{P}}{\alpha_e}>\frac{G}{2\pi}$, and  
(\ref{Lemma6_condition}) holds for 
\begin{equation} \label{variance_condition}
R'>\ln (e G /2)=\ln (G/2) + 1.
\end{equation}

We now state the main result of the paper which will be proven in the following sections:
\begin{prop} \label{main}
The proposed wiretap coding scheme with $\sigma_s^2=P$ achieves strong secrecy for any message distribution $p_{\M}$ (and thus semantic security) for any secrecy rate
$$R < C_b -C_e -\ln\left(2G^2/\pi\right).$$
\end{prop}

\subsection{Secrecy}
The received lattice at Eve's end is $\Hh_e \Lambda$, where $\Hh_e=\diag(\Hh_{e,1},\ldots,\Hh_{e,k})$. Since the message $\M$ and the channel $\Hh_e^k$ are independent, the leakage can be expressed as follows:
{\allowdisplaybreaks
\begin{align*}
&\I(\M;\Z^k,\Hh_e^k)=\I(\M;\Hh_e^k)+\I(\M;\Z^k|\Hh_e)=\I(\M;\Z^k|\Hh_e)=\\
&=\mathbb{E}_{\Hh_e}\left[\I(p_{\M|\Hh_e};p_{\Z^k|\Hh_e})\right]=\mathbb{E}_{\Hh_e}\left[\I(p_{\M};p_{\Z^k|\Hh_e})\right]
\end{align*}
}
We want to show that the \emph{average} leakage with respect to the fading is small. In order to do so, we will show that the output distributions $p_{\Z^k|\Hh_e}$ are close to a Gaussian distribution with high probability. For a fixed realization $H_e=\diag(h_{e,1},\ldots,h_{e,k})$, $H_e \X^k \sim D_{H_e \Lambda_e + H_e \boldsymbol{\lambda}_m, \sqrt{H_e H_e^{\dagger}}\sqrt{P}}$. Using Lemma \ref{extended_Regev_Lemma} with $\Sigma_1=H_e H_e^{\dagger} P$, $\Sigma_2=\sigma_b^2 I$,  
\begin{equation} \label{V}
\mathbb{V}(p_{\Z^k|H_e},f_{\Sigma_0})\leq \epsilon
\end{equation}
provided that 
\begin{equation} \label{fading_secrecy_condition}
\varepsilon_{H_e \Lambda_e}(\sqrt{\Sigma})=\varepsilon_{\sqrt{\Sigma}^{-1} H_e \Lambda_e}(1) \leq \epsilon \leq \frac{1}{2},
\end{equation}
where we define $\Sigma_0=H_e H_e^{\dagger} P + \sigma_b^2 I$, $\Sigma=\frac{(H_e H_e^{\dagger})^{-1}}{P} + \frac{I}{\sigma_b^2}$. 
If (\ref{V}) holds, then it follows from \cite[Lemma 2]{LLBS} that 
\begin{equation} \label{theorem4}
\I(p_{\M};p_{\Z^{k}|H_e}) \leq 8 k\epsilon R  - 8 \epsilon \log 8 \epsilon.
\end{equation}

Recalling the upper bound (\ref{Micciancio_Regev_dual}), we have
\begin{align} 
 & \eta_{\epsilon}(\sqrt{\Sigma^{-1}}H_e \Lambda) \leq 
 \frac{2\sqrt{k}}{\lambda_1(\sqrt{\Sigma}(H_e^{\dagger})^{-1}\Lambda^*)}. \label{faded_smoothing_parameter} 
 \end{align}
Using (\ref{dual_lattice}) and the arithmetic mean - geometric mean inequality, 
{\allowdisplaybreaks
\begin{align*}
&\lambda_1(\sqrt{\Sigma}(H_e^{\dagger})^{-1}\Lambda^*)=2\lambda_1(\sqrt{\Sigma}(H_e^{\dagger})^{-1}\overline{\psi(\mathcal{O}_F^{\vee})})=\\
&=2\min_{x \in \mathcal{O}_F^{\vee} \setminus \{0\}} \norm{\sqrt{\Sigma}(H_e^{\dagger})^{-1}\overline{\psi(x)}} \geq \\   
& \geq 2 \min _{x \in \mathcal{O}_F^{\vee} \setminus \{0\}}  \sqrt{k} \prod_{i=1}^k \Bigg(\frac{P\sigma_e^2}{\sigma_e^2 +P\abs{h_{e,i}}^2}\Bigg)^{\frac{1}{2k}} \prod_{i=1}^k \abs{\sigma_i(x)}^{\frac{1}{k}}= \\  
&=\frac{2\sqrt{k}\sqrt{P}\sigma_e}{G \prod_{i=1}^k (\sigma_e^2+P\abs{h_{e,i}}^2)^{\frac{1}{2k}}}.  
\end{align*}
}
The last equality follows from the fact that 
\begin{align}
&\min _{x \in \mathcal{O}_F^{\vee} \setminus \{0\}} \prod\nolimits_{i=1}^k\abs{\sigma_i(x)}^{\frac{1}{k}}=\min _{a \in \mathcal{O}_F^{\vee} \setminus \{0\}} \abs{N_{K/\Q}(a)}^{\frac{1}{2k}}= \notag\\
&=N(\mathcal{O}_F^{\vee})^{\frac{1}{2k}}=\frac{1}{\abs{d_F}^{1/2k}}=\frac{1}{G}. \label{norm_codifferent}
\end{align}
Replacing in (\ref{faded_smoothing_parameter}), we find that 
for $\epsilon=2^{-2k}$,
$$\eta_{\epsilon}(\sqrt{\Sigma^{-1}}H_e \Lambda) \leq G \prod\nolimits_{i=1}^k (\sigma_e^2 +P \abs{h_{e,i}}^2)^{\frac{1}{2k}}/\sqrt{P}\sigma_e.$$  
Equivalently, in terms of flatness factor we have
$$\varepsilon_{\sqrt{\Sigma^{-1}}H_e \Lambda} \left(\frac{G\prod\nolimits_{i=1}^k (\sigma_e^2+P\abs{h_{e,i}}^2)^{\frac{1}{2k}}}{\sqrt{2\pi P}\sigma_e}\right)\leq 2^{-2k}$$
for fixed fading $H_e$. 
Given $\delta>0$, the law of large numbers (\ref{LLN_Eve}) implies that $\mathbb{P}\left\{\prod_{i=1}^k \left(1+\frac{P}{\sigma_e^2}\abs{h_{e,i}}^2\right)^{\frac{1}{k}}>e^{C_e+\delta}\right\} \to 0$. 
Now suppose that 
\begin{equation} \label{sigma_fading}
\alpha_eG e^{\frac{C_e +\delta}{2}}/\sqrt{2\pi P} \leq 1. 
\end{equation}
We can bound the leakage as follows:
\begin{flalign}
&\mathbb{E}_{\Hh_e}\left[\I(p_{\M};p_{\Z^k|\Hh_e})\right]\leq \notag &\\
&\!\! \leq \mathbb{P}\Big\{\prod\limits_{i=1}^k \Big(1+\frac{P\abs{h_{e,i}}^2}{\sigma_e^2} \Big)^{\frac{1}{k}}>e^{C_e+\delta}\Big\} (kR)+ \notag & \\ 
&\!\!+\mathbb{E}_{\Hh_e}\!\!\left[\I(p_{\M};p_{\Z^k|\Hh_e}) \;\Big|\; \prod\limits_{i=1}^k \Big(1+\frac{P\abs{h_{e,i}}^2}{\sigma_e^2} \Big)^{\frac{1}{k}} \!\!\leq e^{C_e+\delta}\right] \!\!& \label{leakage_sum}
\end{flalign}
The first term vanishes when $k \to \infty$. \\
Now consider the second term. Under the hypothesis that $\prod\nolimits_{i=1}^k \left(1+\frac{P}{\sigma_e^2} \abs{h_{e,i}}^2\right)^{\frac{1}{k}} \leq e^{C_e+\delta}$, we have 
\begin{align*}
& \varepsilon_{\sqrt{\Sigma^{-1}}H_e \Lambda_e}(1)=\varepsilon_{\alpha_e \sqrt{\Sigma^{-1}} H_e \Lambda}(1) \leq \varepsilon_{\sqrt{\Sigma^{-1}}H_e\Lambda}\left(\frac{G e^{\frac{C_e +\delta}{2}}}{\sqrt{2\pi P}}\right) \leq \\
& \leq \varepsilon_{\sqrt{\Sigma^{-1}}H_e\Lambda}\left(\frac{G\prod\nolimits_{i=1}^k (\sigma_e^2+P\abs{h_{e,i}}^2)^{\frac{1}{2k}}}{\sqrt{2\pi P}\sigma_e}\right)\leq 2^{-2k}. 
\end{align*}
Using (\ref{theorem4}), the second term is also vanishing and the lattice coding scheme achieves strong secrecy over Eve's channel. \\
From the conditions (\ref{sigma_fading}) and (\ref{alpha_e}), we find that in order to have strong secrecy we need 
$e G e^{C_e+\delta} \leq e^{R'}$, 
or equivalently 
$R' \geq C_e+\delta + 1+\ln(G)$. 
Since this is true for any $\delta>0$, we find that a rate 
\begin{equation} \label{R_prime}
R' \geq C_e+ 1+\ln(G).
\end{equation}
is required for secrecy. 
\begin{rem}
Although we focused on ergodic fading, the same scheme achieves strong secrecy over the Gaussian and static fading wiretap channels. In fact, for these models 
the first term in (\ref{leakage_sum}) is zero, and the second term still vanishes.   
\end{rem}

\subsection{Reliability}
We suppose that Bob performs MMSE-GDFE preprocessing as in \cite{Damen_ElGamal_Caire}: let $\rho_b=\frac{P}{\sigma_b^2}$, and consider the QR decomposition
$$\widetilde{H}=\left(\begin{array}{c} H \\ \frac{1}{\rho_b}I\end{array}\right)=\left(\begin{array}{c} Q_1 \\ Q_2 \end{array}\right).$$
Observe that $\norm{\mathbf{y}-H_b \mathbf{x}}^2 + \frac{1}{\rho_b} \norm{\mathbf{x}}^2 =\norm{Q_1^{\dagger} \mathbf{y} - R \mathbf{x}}^2+C,$
where $C$ is some constant which does not depend on $\mathbf{x}$. \\
Since the distribution of $\mathbf{x}$ is not uniform, MAP decoding is not equivalent to ML. However, similarly to \cite[Theorem 5]{LLBS}, for fixed $H_b$ which is known at the receiver, the result of MAP decoding can be written as
{\allowdisplaybreaks
\begin{align*}
&\hat{\mathbf{x}}_{\text{MAP}}=\argmax_{\mathbf{x} \in \Lambda_b} p(\mathbf{x} | \mathbf{y})= \argmax_{\mathbf{x} \in \Lambda_b} p(\mathbf{x}) p(\mathbf{y}|\mathbf{x})=\\
&=\argmax_{\mathbf{x} \in \Lambda_b} e^{-\frac{\norm{\mathbf{x}}^2}{2P}}e^{-\frac{\norm{\mathbf{y}-H_b\mathbf{x}}^2}{2\sigma_b^2}}= \\
&=\argmin_{\mb{x} \in \Lambda_b} \left(\frac{1}{\rho_b} \norm{\mb{x}}^2 + \norm{\mb{y} - H_b \mb{x}}^2\right)=
\argmin_{\mb{x} \in \Lambda_b} \norm{Q_1^{\dagger}\mb{y}-R\mb{x}}^2
\end{align*}
}
Thus, Bob can compute $$\mb{y}'=Q_1^{\dagger}\mb{y}=R\mb{x}+\mb{v},$$
where $\mb{v}=Q_1^{\dagger} \mb{w}_b -\frac{1}{\rho_b}(R^{-1})^{\dagger} \mb{x}$ \cite{Damen_ElGamal_Caire}. The noise $\mathbf{v}$ is the sum of a discrete Gaussian 
with distribution $D_{\Lambda',\sqrt{\Sigma_1}}$, where $\Lambda'=\frac{1}{\rho_b}(R^{-1})^{\dagger}\Lambda_b$, $\Sigma_1=\frac{\sigma_b^2}{\rho_b}(RR^{\dagger})^{-1}$, and of a continuous Gaussian random variable $f_{\sqrt{\Sigma_2}}$, where  $\Sigma_2=\sigma_b^2 Q_1Q_1^{\dagger}$. 

For any message $m \in \mathcal{M}$,  $P_e(m)\leq \mathbb{P}\left\{ \mb{v} \notin \mathcal{V}(R\Lambda_b)\right\}$ and consequently the same upper bound holds for the the average:
$$P_e = \sum_{m \in \mathcal{M}} P_e(m) p(m) \leq \mathbb{P}\left\{ \mb{v} \notin \mathcal{V}(R\Lambda_b)\right\}.$$
Although $\mb{v}$ is not Gaussian, we will show that its tails behave similarly to a Gaussian random variable. \\
A random vector $\mb{v}$ taking values in $\C^{k}$ is \emph{$\delta$-subgaussian} with parameter $\sigma$ if $\forall \mb{t} \in \C^{k}$, $\mathbb{E}[e^{\Re(\mb{t}^{\dagger}\mb{v})}] \leq e^{\delta}e^{\frac{\sigma^2}{2}\norm{\mb{t}}^2}$. Note that for a complex Gaussian vector $\mb{z} \sim \mathcal{N}_{\C}(0,\Sigma)$, 
$\mathbb{E}[e^{\Re(\mb{t}^{\dagger}\mb{v})}]=e^{\frac{1}{2} \mb{t}^{\dagger}\Sigma \mb{t}}$. \\
Let's suppose that a fixed message $m$ has been transmitted, so that $\X^k \sim D_{\Lambda_e + \boldsymbol{\lambda}_m,\sqrt{P}}$. The following result holds (see also \cite[Lemma 2.8]{Micciancio_Peikert}).
\begin{lem}
Let $\X^k \sim D_{\Lambda + \mb{c},\sigma}$ be a $k$-dimensional discrete complex Gaussian random variable, and let $A \in M_k(\C)$. Suppose that $\epsilon_{\Lambda}(\sigma)<1$. Then $\forall \mb{t} \in \C^k$,
$$\mathbb{E}[e^{\Re(\mb{t}^{\dagger} A \mb{x})}] \leq \left(\frac{1+\epsilon_{\Lambda}(\sigma)}{1-\epsilon_{\Lambda}(\sigma)}\right) e^{\frac{\sigma^2}{2}\norm{A^{\dagger} \mb{t}}^2}.$$ 
\end{lem}
It follows that $\X^k$ is $\delta$-subgaussian with parameter $\sqrt{P}$ for $\delta=\ln\left(\frac{1+\epsilon}{1-\epsilon}\right)$ provided that 
$\epsilon=\epsilon_{\Lambda_e}(\sqrt{P})<1$,
which is guaranteed by (\ref{variance_condition}). This is weaker than the condition (\ref{R_prime}) we have already imposed for secrecy, so it doesn't affect the achievable 
rate. Consequently, for the equivalent noise $\mathbf{v}$, 
\begin{align*}
&\mathbb{E}[e^{\Re(\mb{t}^{\dagger}\mb{v})}]=\mathbb{E}\left[e^{\Re(\mb{t}^{\dagger} Q_1^{\dagger} \mb{w}_{b})}\right]\mathbb{E}\left[e^{-\Re\left(\frac{1}{\rho_b} \mb{t}^{\dagger} (R^{-1})^{\dagger} \mb{x}\right)}\right] \leq \\
& \leq \left(\frac{1+\epsilon}{1-\epsilon}\right) e^{\frac{\sigma_b^2}{2}\mb{t}^{\dagger}\left(Q_1^{\dagger}Q_1+\frac{1}{\rho_b} (R^{-1})^{\dagger}R^{-1}\right)\mb{t}}= \left(\frac{1+\epsilon}{1-\epsilon}\right) e^{\frac{\sigma_b^2}{2}\norm{t}^2}.
\end{align*}
This implies that the tails of $\mb{v}$ vanish exponentially fast: from \cite[Theorem 2.1]{Hsu_Kakade_Zhang}, it follows that $\forall t>0$,
$$\mathbb{P}\left\{\norm{\mb{v}}^2/k \sigma_b^2> 1 + 2 \sqrt{t/k} + 2 t \right\} \leq e^{\delta} e^{-t}.$$  

In particular, taking 
$\eta=\sqrt{\frac{t}{k}}$, 
we find that $\forall \eta >0$,
$$\mathbb{P}\left\{\norm{\mb{v}}^2/k \sigma_b^2> 1 + \eta\right\} \leq e^{\delta} e^{-k \eta^2}.$$ 

Let $d_R$ denote the minimum distance in the received lattice:
{\allowdisplaybreaks
\begin{align}
&d_R^2=\min_{\lambda \in \Lambda_b \setminus \{0\}} \sum_{i=1}^k \abs{R_i \lambda_i}^2=\min_{\mb{x} \in \psi(\mathcal{O}_F) \setminus \{0\}} \alpha_b^2 \sum_{i=1}^k \abs{R_i x_i}^2 \geq \notag\\
&\geq  \min_{\mb{x} \in \psi(\mathcal{O}_F) \setminus \{0\}} \alpha_b^2 k \prod_{i=1}^k \left(\frac{1}{\rho_b}+\abs{h_{b,i}}^2\right)^{\frac{1}{k}} \prod_{i=1}^k \abs{x_i}^{\frac{1}{k}} \geq \notag \\
& \geq \alpha_b^2 k \prod_{i=1}^k \left(\frac{1}{\rho_b}+\abs{h_{b,i}}^2\right)^{\frac{1}{k}} \label{d_R}.
\end{align}
}
The previous bound follows from the AM-GM inequality and the fact that the minimum non-zero norm of the code is $1$. We use the same argument as in \cite{LV} to bound $P_e$:
given $\eta>0$, 
\begin{align}
&P_e \leq \mathbb{P}\left\{ \mb{v} \notin \mathcal{V}(R\Lambda_b)\right\}\leq \mathbb{P}\left\{ \mb{v} \notin \mathcal{B}(d_R/2)\right\} \leq \notag \\
&\leq \mathbb{P}\left\{\frac{\norm{\mb{v}}^2}{k\sigma_b^2} \geq 1+\eta\right\}+\mathbb{P}\left\{\frac{d_R^2}{4k\sigma_b^2}<1+\eta\right\}.  \label{error_probability} 
\end{align} 
Since the first term vanishes exponentially fast when $k \to \infty$, we can focus on the second term. From (\ref{d_R}), 
the second term in (\ref{error_probability}) is upper bounded by 
{\allowdisplaybreaks
\begin{align*}
&\mathbb{P}\left\{ \frac{\alpha_b^2}{4\sigma_b^2} \prod_{i=1}^k \left(\frac{1}{\rho_b}+\abs{h_{b,i}}^2\right)^{\frac{1}{k}} < 1+\eta\right\}=\\ 
&=\mathbb{P}\left\{ \frac{1}{k} \sum_{i=1}^k \ln \left(1+\rho_b\abs{h_{b,i}}^2\right) < \ln \left(\frac{4 P(1+\eta)}{\alpha_b^2}\right)\right\}=\\ 
&=\mathbb{P}\left\{ \frac{1}{k} \sum_{i=1}^k \ln \left(1+\rho_b\abs{h_{b,i}}^2\right) < \ln \left(\frac{2G e^{R_b}(1+\eta)}{\pi e}\right)\right\},
\end{align*}
}%
recalling that 
$\alpha_b^2 \approx \frac{2\pi e P}{G e^{R_b}}$ 
from (\ref{alpha_e}) and the fact that $\abs{\Lambda_b/\Lambda_e}=e^{kR}$. Since the left hand side tends to $C_b$ when $k \to \infty$ due to the law of large numbers (\ref{LLN_Bob}), the last expression will vanish provided that $R_b < C_b- \ln \left(\frac{2G}{\pi e}\right)-\ln(1+\eta)$.
Since $\eta$ is arbitrary, any rate 
\begin{equation} \label{R_b}
R_b=R+ R' < C_b- \ln \left(2G/\pi e\right)
\end{equation}
is achievable for Bob. 
From equations (\ref{R_prime}) and (\ref{R_b}), the proposed coding scheme achieves strong secrecy for any message distribution (and thus semantic security) for any secrecy rate
$$R < C_b -C_e -\ln\left(2G^2/\pi\right).$$
This concludes the proof of Proposition \ref{main}. 

\section*{Acknowledgements}
Cong Ling's work was supported in part by FP7 project PHYLAWS (EU FP7-ICT 317562). The research of R. Vehkalahti was funded by Finnish Cultural Foundation.\\
The authors would like to thank Jean-Claude Belfiore and Hamed Mirghasemi for useful discussions.

\begin{footnotesize}

\end{footnotesize}

\end{document}